\documentclass[aps,pre,twocolumn,groupedaddress]{revtex4}

\usepackage{graphicx}
\usepackage{amssymb}
\usepackage{bm}

\begin{document}


\title{Shear strength properties of wet granular materials}


\author{Vincent Richefeu}
\email[]{richefeu@lmgc.univ-montp2.fr}
\author{Moulay Sa\"{i}d El Youssoufi}

\author{Farhang Radja\"{i}}

\affiliation{Laboratoire de M\'{e}canique et G\'{e}nie Civil UMR CNRS 5508, Cc. 048, Universit\'{e} Montpellier 2,\\ 
Place Eug\`{e}ne Bataillon, 34095 Montpellier Cedex 5, France}

\date{\today}

\begin{abstract}

We investigate shear strength properties of wet granular materials in the pendular state (i.e. the state where the liquid phase is discontinuous) as a function of water content. Sand and glass beads were wetted and tested in a direct shear cell and under various confining pressures. In parallel, we carried out three-dimensional molecular dynamics simulations by using an explicit equation expressing capillary force as a function of interparticle distance, water bridge volume and surface tension. We show that, due to the peculiar features of capillary interactions, the major influence of water content over the shear strength stems from  the  distribution of liquid bonds. This property results in shear strength saturation as a function of water content. We arrive at the same conclusion by a microscopic analysis of the shear strength. We propose a model that accounts for the capillary force, the granular texture and particle size polydispersity. We find fairly good agreement of the  theoretical estimate of the shear strength with both experimental data and simulations. From numerical data, we analyze the connectivity and anisotropy of different classes of liquid bonds according to the sign and level of the normal force as well as the bond direction. We find that weak compressive bonds are almost isotropically distributed whereas strong compressive and tensile bonds have a pronounced anisotropy. The probability distribution function of normal forces is exponentially decreasing for strong compressive bonds, a decreasing power-law function over nearly one decade for weak compressive bonds and an increasing linear function in the range of tensile bonds. These features suggest that different bond classes do not play the same role with respect to the shear strength.   

\end{abstract}


\maketitle



\section{Introduction}

Capillary cohesion is known to influence strongly the strength and flow properties of granular materials. For example, sandcastles keep standing basically due to small amounts of water between sand grains \cite{hornbaker97,halsey98}. At low levels of water content,  the water forms a discontinuous phase composed of interparticle bridges that are unevenly distributed in the bulk (the pendular state). Obviously, cohesion effects appear only at  low confining pressures, e.g. in surface soils. It is a common observation that, when plowing a wet granular soil, large cohesive aggregates are formed. The largest capillary cohesion force for millimeter-size sand grains is about $4 \times 10^{-4}$ N independently of meniscus volume. This force is nearly four times the grain weight, allowing thus for the formation of cohesive aggregates. Transformations involving primary particle agglomeration into coherent granules are of special interest in many applications in a wide range of industries such as pharmaceuticals, agronomic products and detergents \cite{Bika01,Iveson02}.

Although capillary phenomena at the interface between two solid bodies are well understood, it is much less clear how a collection of grains reacts to the presence of a liquid. The issue is basically the same as in cohesionless granular media where the influence of interparticle friction on the shear strength depends both on the features of the friction law itself and the granular structure. Similarly, the question here is the extent to which the features of cohesion interactions are reflected in a global property such as shear strength. At least two factors seem to be important:  the local force thresholds and the distribution of cohesive bonds in the bulk \cite{halsey98,Kohonen04}. There are reasons to think that the spatial distribution of water bonds should prevail over the absolute amount of water going to each bond. Our results, as we shall see in this paper, credit this point.

There is a vast literature dealing with the experimental behavior of wet granular media. In soil mechanics, where stability considerations are of primary importance, the influence of water content on unsaturated earth is mostly studied in shear tests through the Coulomb cohesion parameter representing the shear strength at zero confining stress \cite{mitchell05}. On the other hand, the flowability of wet powders is expressed in terms of tensile strength measured as the fluidization threshold under vibration or air flow \cite{Castellanos02}. Direct measurements of tensile strength have been reported more recently \cite{Pierrat97,Betz02,Kim03}. In granular media, it is generally much more difficult to access local information such as contact forces or liquid bonds. Few investigations have recently been reported to visualize liquid bonds by means of the index matching technique \cite{Fournier05,Kohonen04}. These observations underline the influence of the distribution of water bonds.

A detailed description of the behavior at the particle scale can be obtained by means of molecular dynamics (MD) simulations with an appropriate prescription of force laws. Recently, several simulations of wet granular media have been reported \cite{Thornton04,Lian98,Rhodes01}. 
Mikami et al. used this type of simulation together with a regression expression for the liquid bridge force as a function of  liquid bridge volume and separation distance between particles \cite{Mikami98}. They mainly studied bubbling behavior and agglomerate formation in a fluidized bed and they found realistic results. Dense agglomerates were simulated by Gr\"{o}ger et al. using a cohesive discrete element method \cite{Groger03}. They found a good agreement with experimental data for the yield stress at all confining pressures down to the value of the tensile stress. Shear strength behavior of unsaturated granulates was also studied numerically by Jiang et al. as a function of suction (pressure difference between liquid and gas) \cite{Jiang04}.

In this paper, we investigate the cohesive behavior of wet dense granular packings under monotonous shearing by means of experiments and three-dimensional MD simulations. We also analyze the shear strength from a microscopic expression of the stress tensor. The experiments are described in section \ref{sec:exp}. We study the Coulomb cohesion as a function of water content for four different materials. The simulations are presented in section \ref{sec:simu}. We use an explicit expression of the capillary force as a function of interparticle distance, bridge volume and surface tension. The Coulomb cohesion is studied as a function of water content. We also investigate the bond connectivity and force distributions. In section \ref{sec:analysis}, we propose a new expression for the shear strength that accounts for particle polydispersity as well as material and structural parameters. We finally compare the experimental and numerical results to the theoretical predictions.

\section{\label{sec:exp}Experiments}

The experiments  were designed to measure the shear strength at  low confining pressures ($< 1$ kPa). The principles of the device are similar to those used in several other investigations  \cite{Hubbert51,Schellart00,Rossi03,Mechelen04}. We present here the setup, the materials, the wetting protocol and our main results.   

\subsection{Experimental setup}

\begin{figure}
\includegraphics[width=8.5cm]{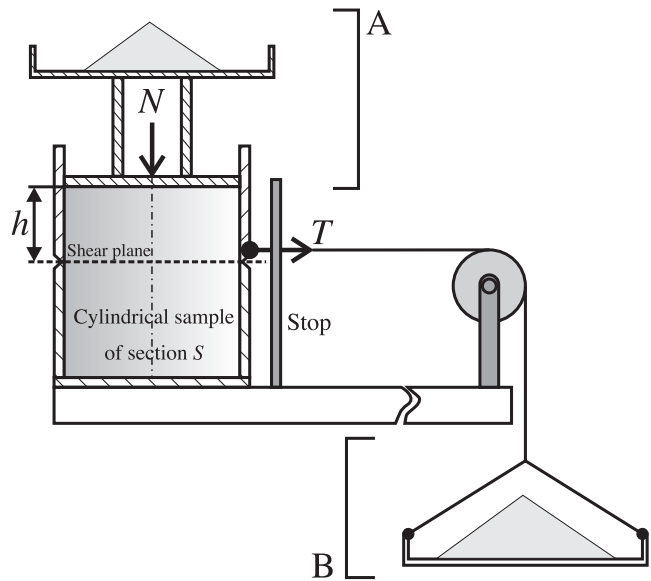}
\caption{Testing cell and shearing setup.\label{fig1}}
\end{figure}

A schematic representation of the shearing setup is shown in Fig.~\ref{fig1}. The wet grains are poured in a plexiglas cylindrical cell and confined by means of a circular lid of area $S$ placed on top of the material. The lid is equipped with a reservoir allowing to impose an overload by adding desired amount of sand. The total vertical force $N$ acting on the sample is the sum of the weights of lid and sand (shown by A on the figure). The cell is composed of two disjoint parts kept together during sample preparation. The upper part can move horizontally with respect to the lower part by pulling on a rope attached to it and which supports a cupel through a pulley (shown by B on the figure). The pulling force $T$ can be increased by adding sand into the cupel. The friction force between the two parts of the cell is reduced by water lubricating the rims and we checked that it remains negligibly small during shear. In order to reduce the friction force exerted by the material along the walls (Janssen effect), the thickness $h$ of the upper part of the sample was taken to be below the diameter of the cell ($46$ mm). 
The heights of the upper and lower parts  are about $10$ mm and $15$ mm respectively.
The sample is sheared along the common section of the two parts of the cell. This shear plane is subjected to a tangential stress $\tau = T/S$ and a normal stress $\sigma = N/S + \rho g h$, where $\rho$ is the bulk density and $g$ is the gravity.

In the experiments, we gradually increase the shear stress $\tau$ for a fixed value of $\sigma$. Unstable shearing occurs when $\tau$ reaches the shear strength $\tau_m$, resulting in a sudden slide of the upper part of the sample. 
The upper part is stopped by collision with two bars located $5$ mm away from the cell.
We did not measure the displacements. We recorded $\tau_m$ for different values of $\sigma$ in the range varying from $200$ Pa to $800$ Pa, and for different values of water content.     

\subsection {Materials and wetting protocol}

We used four types of materials: (1) a sand composed of angular grains with diameter in the range from $0.1$~mm to  $0.4$~mm, (2) ``tightly-graded'' polydisperse glass beads with diameters from $0.4$~mm to $0.5$~mm, (3) ``well-graded'' polydisperse glass beads with diameters from $0.4$~mm to $0.8$~mm, and (4) monodisperse glass beads of diameter $1$~mm.

In order to wet the grains, we add distilled water to dry material placed in a vessel which is then closed and energetically shaken for about one minute. 
The vessel used for mixing is transparent so that during shaking we can check visually whether the water is homogeneously mixed with the grains. In particular, we continue shaking until all visible water clusters disappear. Increasing the duration of shaking beyond 1 minute did not change the measured values of the shear strength. After mixing, the wetted material is poured into the testing cell. The water content is evaluated by comparing the masses of a sample of the material before and after testing by means of a heat chamber used for drying the sample at $105^\circ$C.
The water content is given by $w = m_w/m_s$, where $m_w$ and $m_s$ are the masses of water and grains, respectively. The wet materials were tested for water contents below $0.05$ corresponding to the pendular state for our materials.
The experiments were performed at ambient conditions. Each experiment lasted a few minutes. The loss of liquid was always below two percent. This loss is not only due to evaporation but also due to partial wetting of the internal walls of the cell. It is small enough to assume a constant liquid volume (as in simulations, see below).

\subsection {Results}
 
 Figure \ref{fig2}(a) shows the yield loci $\tau$-$\sigma$ for the sand at several levels of water content $w$. Within experimental precision, the data are well fitted by a straight line, in agreement with the Mohr-Coulomb model
\begin{equation} \label{eq:Mohr-Coulomb}
\tau = \mu \sigma + c,
\end{equation}
where $\mu = \tan \varphi$ is the internal coefficient of friction and $c$ is the Coulomb cohesion. It is remarkable that $\varphi \simeq 33^\circ$ is almost independant of $w$. On the other hand, the Coulomb cohesion increases nonlinearly with $w$ and saturates to $c_m \simeq  600$~Pa at $w_m \simeq 0.03$, as shown in Fig.~\ref{fig2}(b).

\begin{figure}
\includegraphics[height=12cm]{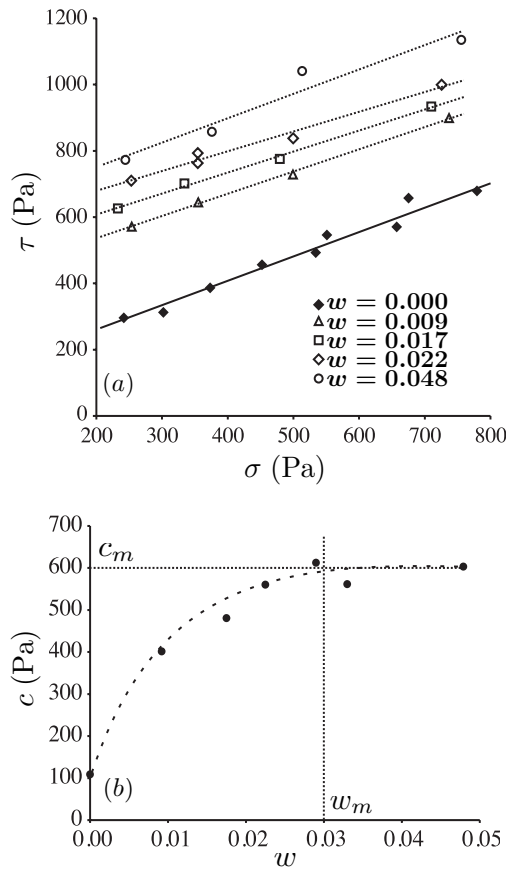}
\caption{(a) Yield loci $\tau$-$\sigma$ of sand for increasing level of water content; (b) the Coulomb cohesion as a function of water content. The dashed line is drawn as a guide to the eyes.\label{fig2}}
\end{figure}

\begin{figure}
\includegraphics[height=12cm]{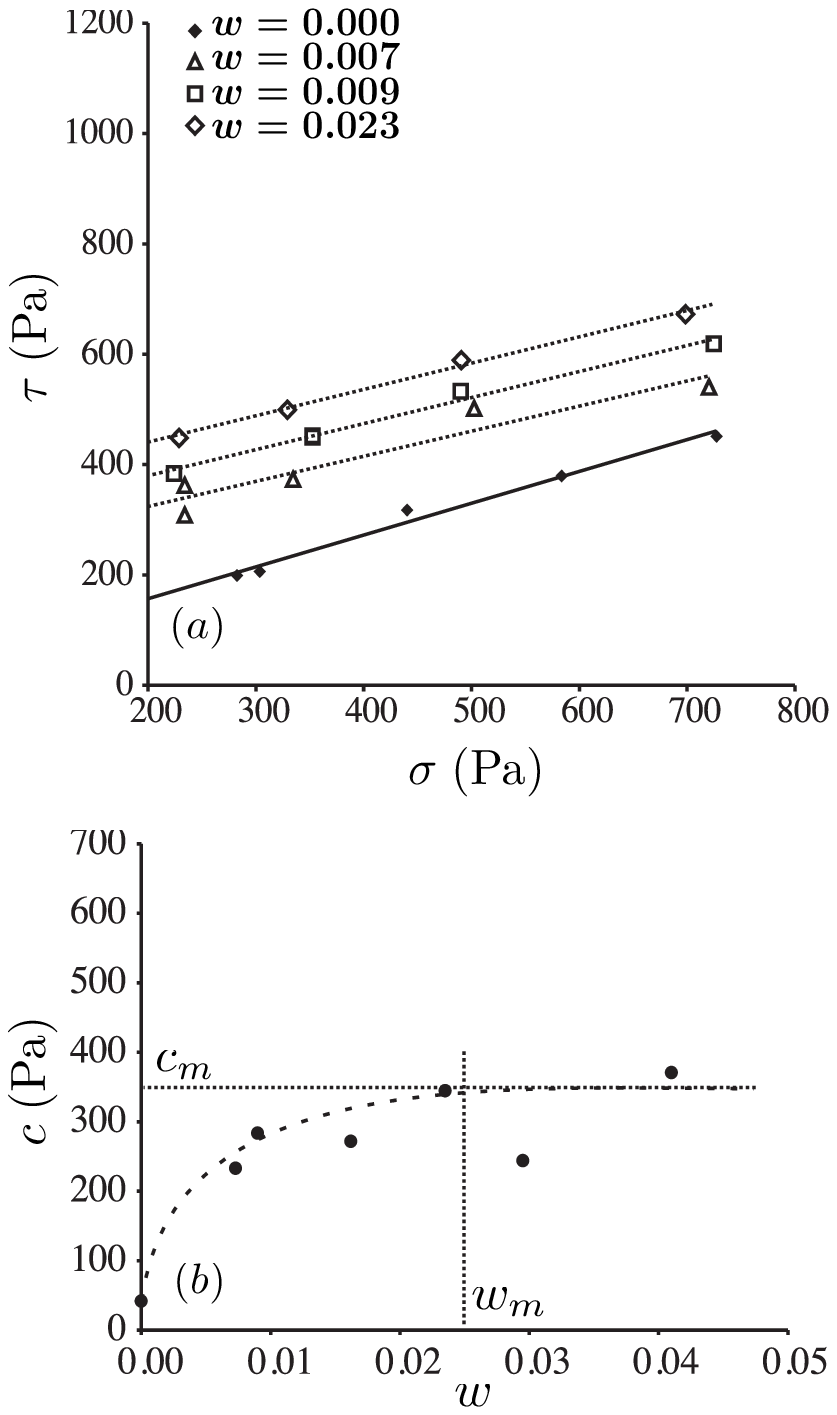}
\caption{(a) Yield loci $\tau$-$\sigma$ of tightly-graded glass beads for increasing level of water content; (b) the Coulomb cohesion as a function of water content.\label{fig3}}
\end{figure}

\begin{figure}
\includegraphics[height=12cm]{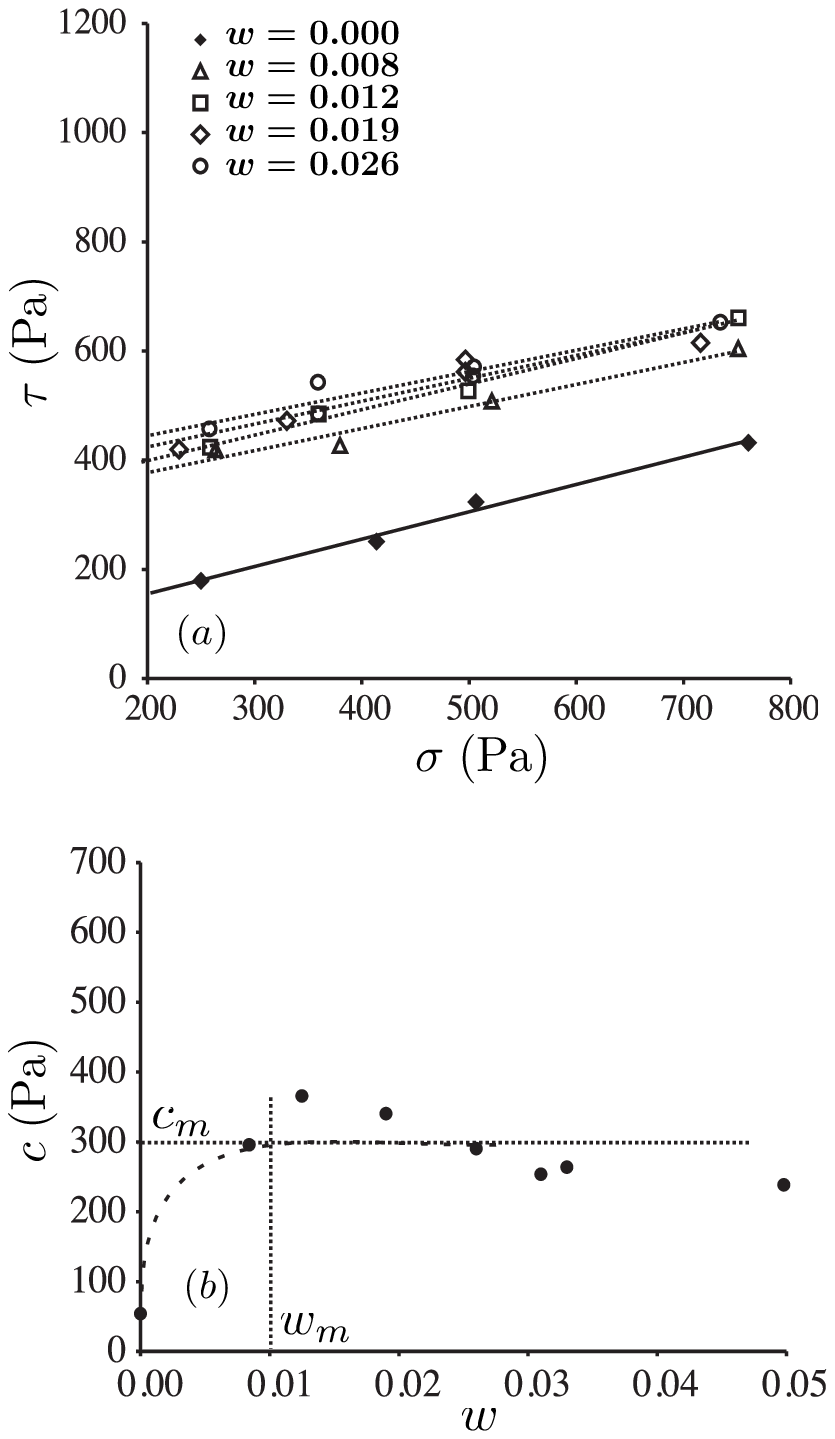}
\caption{(a) Yield loci $\tau$-$\sigma$ of well-graded glass beads for increasing level of water content; (b) the Coulomb cohesion as a function of water content.\label{fig4}}
\end{figure}

\begin{figure}
\includegraphics[height=12cm]{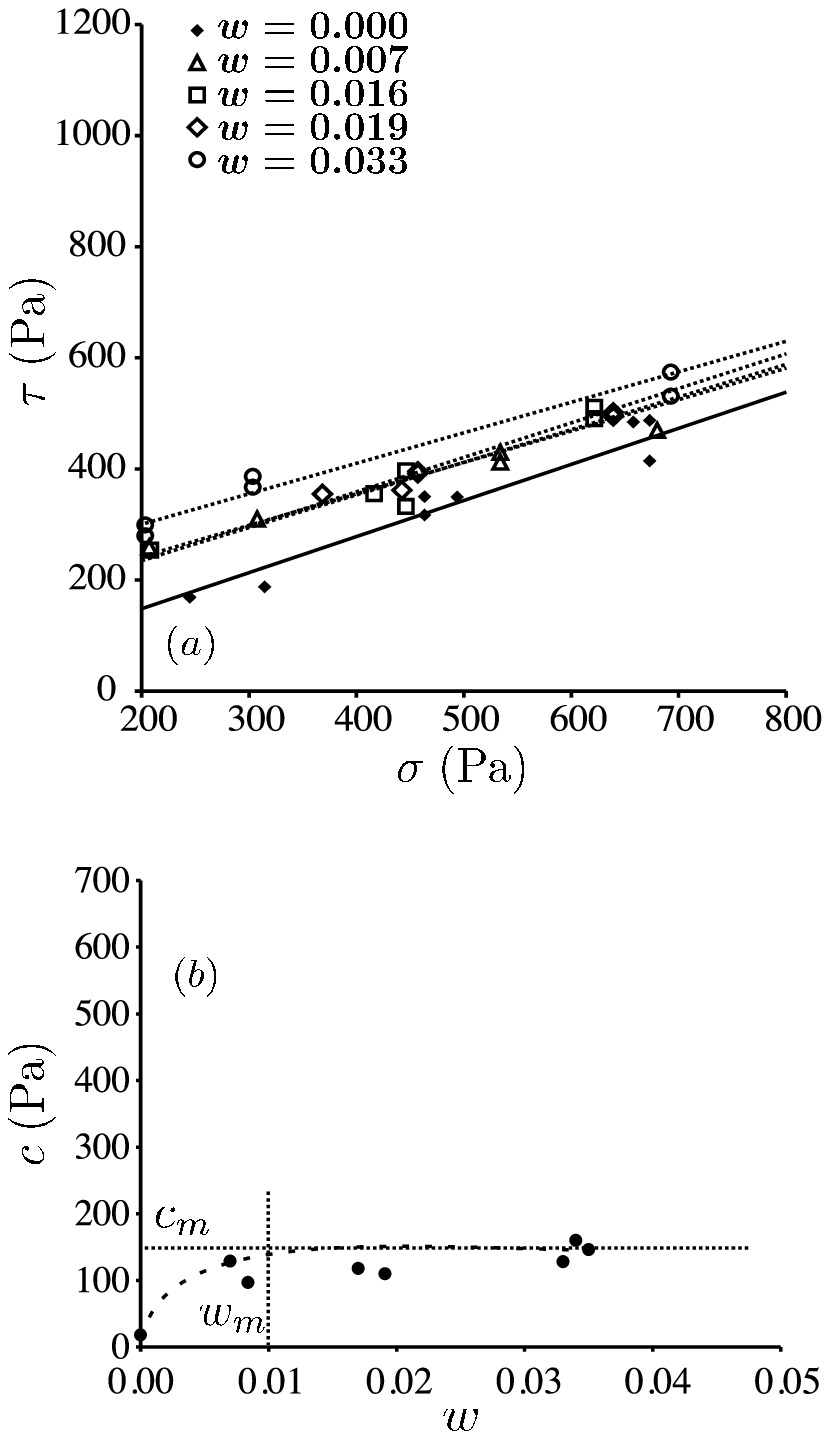}
\caption{(a) Yield loci $\tau$-$\sigma$ of monodisperse glass beads for increasing level of water content; (b) the Coulomb cohesion as a function of water content.\label{fig5}}
\end{figure}

For tightly-graded polydisperse beads we get a similar behavior with $\varphi \simeq 30^\circ$; Figs. \ref{fig3}(a) and (b). However, saturation occurs at a lower level of water content ($w_m \simeq 0.025$) and cohesion ($c_m \simeq 350$~Pa). Figures \ref{fig4}(a) and (b) show the results for well-graded polydisperse beads. Saturation occurs at about $w_m \simeq 0.01$ and $c_m \simeq 300$~Pa. Enhance fluctuations observed in the data from glass beads, compared to the sand, may be attributed to a lower level of cohesion and a tighter size distribution of glass beads. 
In fact, a lower level of cohesion leads to larger mobility of the particles and the porosity increases as the particle sizes are less widely distributed. 
In the case of monodisperse beads, the cohesion jumps from zero for the dry material to a nonzero value ($c_m \simeq 150$~Pa) independently from water content; Figs.~\ref{fig5}(a) and (b). Since we have few data points between $w = 0$ and $w \simeq 0.01$, the saturation level $w_m$ should be below $0.01$. 

Notice that the shear tests provide quite reproducible results. We see that, in Figs.~\ref{fig2}(a)-\ref{fig5}(a), the Mohr-Coulomb line passes through most of data points. The very weak dispersion of the data points about this line shows the high reproducibility of the testing procedure.  This means that the experimental measurement of the Coulomb cohesion is reliable. The different values of $c_m$ for different materials will be discussed in section \ref{sec:analysis} (see also Table~\ref{tab2}). 
 
The value of $w_m$ is less clearly defined and it is likely to depend on two factors: (1) the surface state of the particles  and (2) possible clustering of the liquid phase \cite{halsey98,Kohonen04,Fournier05}. The sand grains have a rough surface requiring more water to form a meniscus than glass beads which are much more smooth. On the other hand, partial clustering of water may occur and this might require a larger amount of water for the formation of liquid bridges although we observed no clustering at the visible parts of the packing through the transparent walls of the testing cell. It is worth noting that it is not straightforward to evaluate the variability of $w$ since each value results from several experiments. If the evaluation is based only on the measurement of the water content at the beginning and at the end of each test, the error would be as small as 2\% and this can not be shown in the figures (it would be smaller than the size of the data point symbols).
 
\section{\label{sec:simu}Simulations}

For the simulations, we employed the framework of the molecular dynamics method \cite{cundall79,allen87}. This method is referred to as Distinct Element Method (DEM) in the geotechnical context. The heart of our simulations is, however, the model of capillary cohesion and its implementation, as well as the way liquid bridges are numerically distributed in the packing. These aspects are detailed below.

\subsection{Molecular Dynamics}

We implemented the basic molecular dynamics method for spherical particles. The equations of motion are integrated according to the velocity Verlet scheme. The normal force between two particles is the sum of a repulsive force as a linear function of the overlap and an attraction force due to the presence of a liquid bridge at contact or for a gap up to a rupture distance (see below). As usual in molecular dynamics simulations, normal dissipation is accounted for by viscous damping. The particles interact also through a Coulomb friction law with a viscous regularization at low sliding velocities. The material is sheared quasistatically in a direct shear cell in the presence of gravity and confining stresses. The damping parameter and the normal stiffness (force per unit overlap) were adjusted in order to get largest time step ($10^{-6}$ s) and small overlaps within numerical stability. The normal stiffness and the interparticle coefficient of friction were $10^3$ N/m and $0.4$, respectively.  

\subsection{Capillary cohesion law}

\begin{figure}
 \includegraphics[width=8cm]{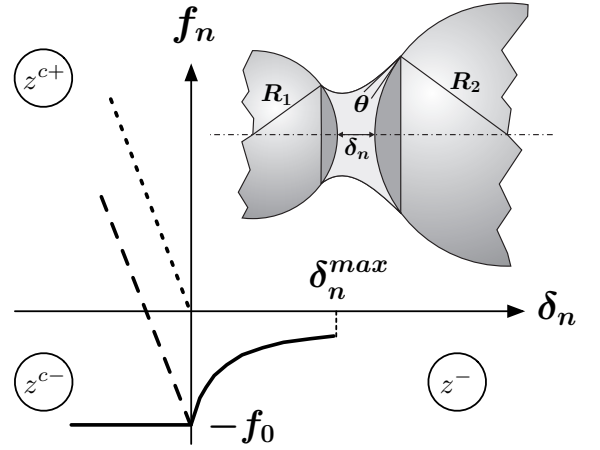}
\caption{Typical behavior of the capillary force $f_n^c$ as a function of the gap $\delta_n$ (solid line).  The elastic repulsive normal force as a result of overlapping ($\delta_n<0$) is shown (short-dashed line). The resultant normal force in this range is shown as well (long-dashed line). $z^{c+}$, $z^{c-}$ and $z^-$ are the partial wet coordination numbers in this range. Inset: Geometry of a liquid bridge between two particles.\label{fig6-7}}
\end{figure}

The capillary attraction force between two particles is a consequence of the liquid surface tension and the pressure difference between liquid and gas phases \cite{Lian93}. For efficient numerical calculation, we need an explicit expression of the capillary force $f_n^{c}$ as a function of the interparticle gap $\delta_n$. By extending the work of Mikami et al. \cite{Mikami98}, it was recently shown by Souli\'e et al. \cite{Soulie05c} that the capillary force can be cast in the following form: 
\begin{equation} \label{eq:cohesion-law}
f_n^{c} =
\left \{
\begin{array}{ll}
- \pi \gamma_s \sqrt{R_1 R_2}\{\exp (A \delta_n^\star +B)+C\} & {\rm for}\; \delta_n^\star > 0 \\
 & \\
- \pi \gamma_s \sqrt{R_1 R_2}\{\exp (B)+C\} & {\rm for} \; \delta_n^\star \leq 0
\end{array}
\right. ,
\end{equation} 
where $R_1$ and $R_2$ are the sphere radii ($R_1 \leq R_2$), $\gamma_s$ is the liquid surface tension, $\delta_n^{\star}=\delta_n / R_2$ (see Fig.~\ref{fig6-7}). The parameters $A$, $B$ and $C$ are functions of the liquid volume $V_b$ of the bond and the contact angle $\theta$ as follows
\begin{equation}
\begin{array}{rcl}
A & = & -1.1 ( V_b^{\star} )^{-0.53},\\
 & & \\
B & = & \left ( -0.148 \ln V_b^{\star} -0.96\right ) \theta^2 \\
 & & - 0.0082 \ln V_b^{\star}  + 0.48,\\
  & & \\
C & = & 0.0018 \ln V_b^{\star} + 0.078,\\
\end{array}
\end{equation}
where $V_b^{\star} = V_b / R_2^3$.

It can be shown that a liquid bond is stable as long as the gap is below a de-bonding distance $\delta_n^{max}$ given by \cite{Lian98}
\begin{equation}
\delta_n^{max} = (1+ 0.5 \theta)V_b^{1/3}.
\end{equation}

\begin{figure}
 \includegraphics[width=7cm]{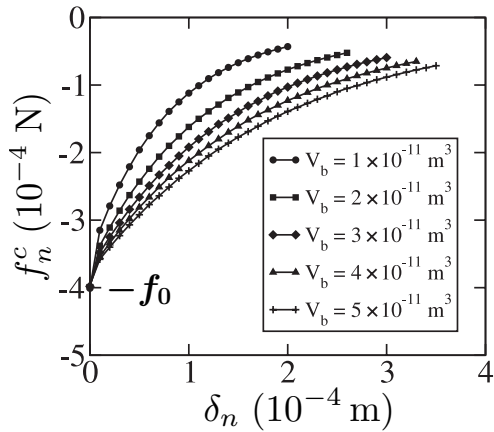}
\caption{Capillary force as a function of the gap according to Eq. \ref{eq:cohesion-law} for increasing liquid volume $V_b$.\label{fig8}}
\end{figure}

Figure \ref{fig6-7} shows a schematic representation of the capillary force as a function of the gap. In Fig.~\ref{fig8} the capillary force $f_n^c$ is displayed as a function of the gap $\delta_n$ according to Eq.~\ref{eq:cohesion-law} for different values of the liquid bridge volume $V_b$. These plots are in perfect agreement with those obtained by other authors by integration of the Laplace-Young equation and verified experimentally \cite{Willett00}.
The largest absolute value $f_0$ of the capillary force occurs at $\delta_n = 0$. It is remarkable that $f_0$ is directly proportional to $\sqrt{R_1 R_2}$ and only very weakly dependent on the liquid volume.
This property, which might seem counter-intuitive, is important for the model that will be introduced in section \ref{sec:analysis}. Hence, with a good approximation, we may write  
%
\begin{equation} \label{eq:kappa}
f_0 = \kappa \sqrt{R_1 R_2},
\end{equation}
where $\kappa$ is a function only of the surface tension and the contact angle. In our case, with glass beads and water bridges, we have $\kappa = 0.4$ N/m. Let us underline here the fact that the capillary bond in the range $\delta_n > 0$ is unstable with respect to the forces acting on two particles. In other words, when pulling two particles apart from one another, the liquid bond fails at zero gap for $f_n = -f_0$. In our simulations, we find that the fraction of liquid bonds in the range $\delta_n > 0$ is always below $15\%$ (see Fig. \ref{fig12}). This shows that the capillary failure threshold $f_0$ is far more important for the failure of a wet material than the de-bonding distance $\delta_n^{max}$. This point will be discussed in more detail in section \ref{sec:analysis}.

\subsection{Sample preparation}

The numerical samples are composed of spherical particles of three different diameters ($2$ mm, $1.5$ mm and $1$ mm) placed randomly in a cylindrical cell in appropriate proportions ($50\%$, $30\%$ and $20\%$) to represent one of our experimental samples composed of glass beads. The initial configuration is prepared under gravity without introducing capillary bonds. Then, we attribute a capillary bond to each pair of particles within the de-bonding distance. Finally, the sample is consolidated under the action of a vertical confining pressure with a zero coefficient of friction. The consolidation is stopped and the coefficient of friction set to $0.4$ as soon as the solid fraction $\phi=0.6$ is reached. The subsequent compaction is negligibly small.  

The volume $V_b$ attributed to a capillary bond between two particles is taken to be proportional to the particle diameters and the intercenter distance, and such that the total volume of all liquid bonds in the sample is equal to that of the added water. Since all particle pairs within the de-bonding distance are considered, the liquid coordination number $z$ (i.e. the average number of liquid bonds per particle) obtained by this procedure has the highest possible value. For our sample we get $z=8$ for $w=0.01$ (see Fig.~\ref{fig16}).
In our simulations, the liquid bond volumes vary by a factor 8 from the contact between the smallest particles to that between the largest particles.  On the 
other hand, the contact angle with a good approximation was set to
zero.  Moreover, since the de-bonding length varies as $V_b^{1/3}$, there is a factor 2 between the shortest and longest de-bonding distances.

 During shearing, the number of liquid bonds evolves and the available liquid must be redistributed in the system. We used two different methods for redistribution: (1) we simply apply the above procedure every time the contact list is updated; (2) the volume of a broken liquid bond is split between the corresponding particles (proportionally to their diameters) and conserved  for the formation of new liquid bonds when a contact occurs with the same particles. 
In this method, the volume of free liquid left after de-bonding is kept with the two particles (and not distributed to the other bonds of the same particles) and used only if a new contact is formed. This implies that, if the initial liquid distribution is homogeneous, then it will remain so during deformation as in the first method. In other words, the liquid will not migrate considerably and one should expect quite similar results from both methods.  
Indeed, in different tests, we found that both methods lead to nearly identical results. Unless mentioned explicitly, all results presented in this section were obtained by the first method. 
  
\subsection{Boundary conditions and driving}

As in experiments, the cylindrical cell is composed of two disjoint parts. The lower part is fixed whereas the upper part moves horizontally, giving rise to a shear plane along the common section of the two parts. We apply a constant vertical load $\sigma$, the same as in experiments, on top of the sample. However, in contrast to experiments, shearing is controlled by imposing a constant horizontal velocity on the upper part. The numerical sample has exactly the same dimensions as in experiments.

\subsection{Results}

\begin{figure}
\includegraphics[width=7cm]{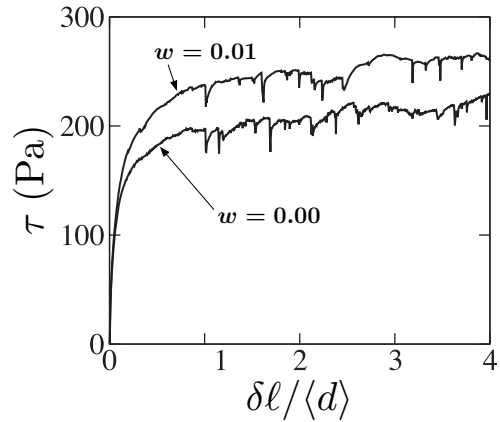}
\caption{The shear stress $\tau$ as a function of shearing distance $\delta \ell$ normalized by the average particle diameter $\langle d \rangle$ for a dry and a wet sample simulated by the molecular dynamics method. The confining stress is $\sigma = 300$ Pa.\label{fig9}}
\end{figure} 

Figure \ref{fig9} shows the stress-strain plot for a dry and a wet sample with $w=0.01$. The initial configuration is the same in both simulations. The initial elastic increase of $\tau$ (up to $\sim 75$~Pa) as a function of $\delta \ell$ is common between the two samples. We observe no stress peak in the wet case. The steady state (``critical state'' in soils mechanics) is reached at $\delta \ell \simeq \langle d \rangle$ for all water contents. The steady state deformation involves numerous instabilities that occur throughout the system and appear in the form of rapid stress drops on the stress-strain plots. We see that in transition from dry to wet materials the frequency of such instabilities declines while their amplitudes grow.

\begin{figure}
\includegraphics[width=7cm]{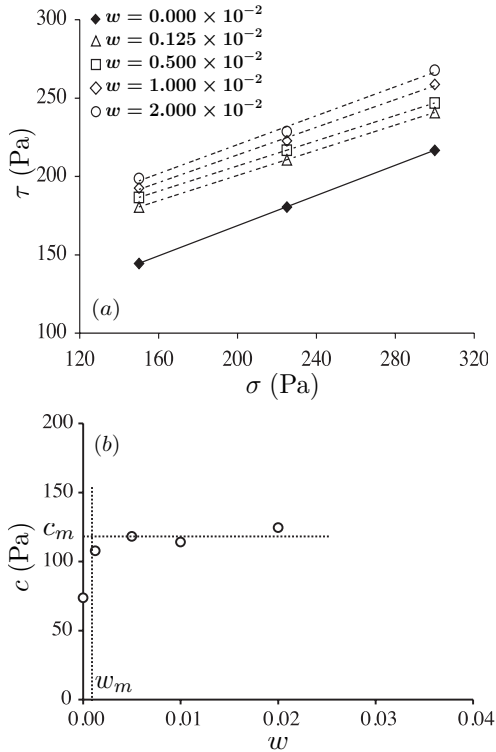}
\caption{(a) Yield loci $\tau$-$\sigma$ from 15 simulations for increasing level of water content; (b) the corresponding values of the Coulomb cohesion $c$ as a function of water content.\label{fig10}}
\end{figure}

We now turn to the evolution of the Coulomb cohesion as a function of $w$. Figure~\ref{fig10}(a) shows fitted yield loci from 15 simulations involving three different values of the confining pressure $\sigma$ and five different values of the water content $w$. The Coulomb cohesion $c$ is drawn as a function of $w$ in Fig.~\ref{fig10}(b). The latter is very similar to the corresponding experimental plot (Fig.~\ref{fig5}(b)) for monodisperse glass beads. We observe a saturation of $c$ at still lower levels of water content  ($w_m \simeq 0.001$). Note also that, while the average grain size is nearly the same in these simulations and in the case of monodisperse experimental glass beads, the maximum cohesion $c_m=120$ Pa in the simulations is very close to that ($150$ Pa) for $1$ mm glass beads.

In contrast to the experiments, where the stresses are measured at the walls, it is also possible to compute the stress tensor for grain-to-grain forces in the simulations (see below). However, we found that in our simulations, the results from these two methods do not coincide. This is because in direct shearing, wall effects give rise to large stress gradients in the bulk. This point has been analyzed in detail by Thornton and Zhang \cite{Thornton01}. We used here the wall stresses $\tau$ and $\sigma$ for comparison with the corresponding experimental values.  

\begin{figure}
\includegraphics[width=7cm]{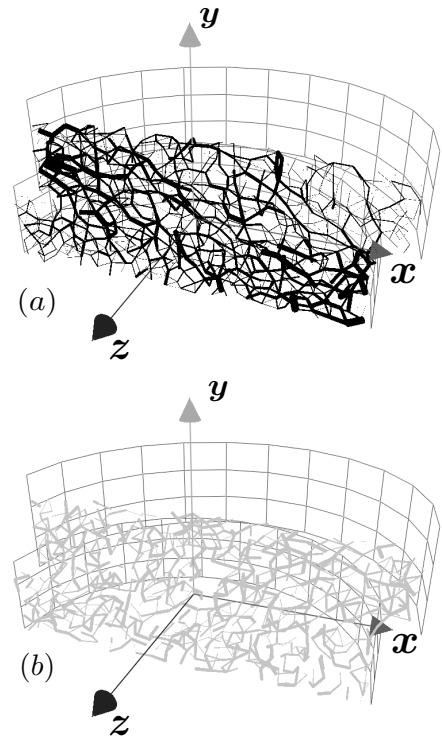}
\caption{(Color online) Compressive force network (a) and tensile force network (b) in a thin vertical layer. Weak forces are not shown. $xz$ is the shear plane and $y$-axis points upward.\label{fig11}}
\end{figure}

\begin{figure}
\includegraphics[width=7cm]{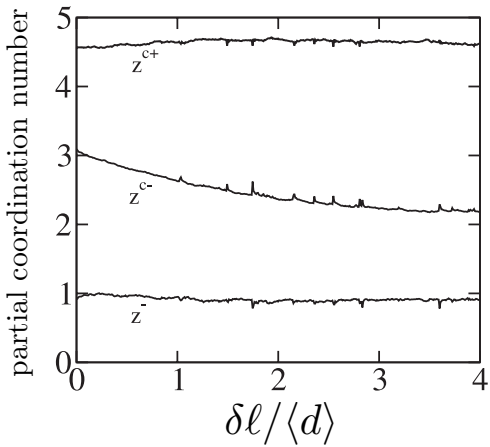}
\caption{Partial coordination numbers as a function of shearing distance.\label{fig12}}
\end{figure}

Figure \ref{fig11} displays a typical example of the force network in a thin vertical layer (parallel to $xy$ plane). We observe a strongly inhomogeneous transmission of both compressive and tensile forces.   
The  liquid bonds belong to three different classes: (1) contacts carrying a compressive (positive) force; (2) contacts carrying a tensile (negative) force; (3) liquid bonds with no contact and thus carrying a tensile force. We denote the corresponding partial coordination numbers by $z^{c+}$, $z^{c-}$ and $z^{-}$, respectively (see Fig.~\ref{fig6-7}). Fig.~\ref{fig12} shows the evolution of these partial coordination numbers as a function of shear displacement. Interestingly, although the bonds are optimally distributed in the sample (according to the first redistribution method), $z^-$ is only about one bond per particle. This means that the overall contribution of this class, involving a low number of bonds and rather weak forces, to stress transmission is marginal. The evolution of particle connectivity is mainly reflected in the regular fall-off of $z^{c-}$ during shear. This class, with numerous contacts and a high force level, provides the largest contribution to the transmission of tensile stresses in the packing. The rather large and constant value of $z^{c+}$ is a reflection of the fact that the packing is globally subjected to boundary compressive stresses.    
 
\begin{figure}
\includegraphics[width=7cm]{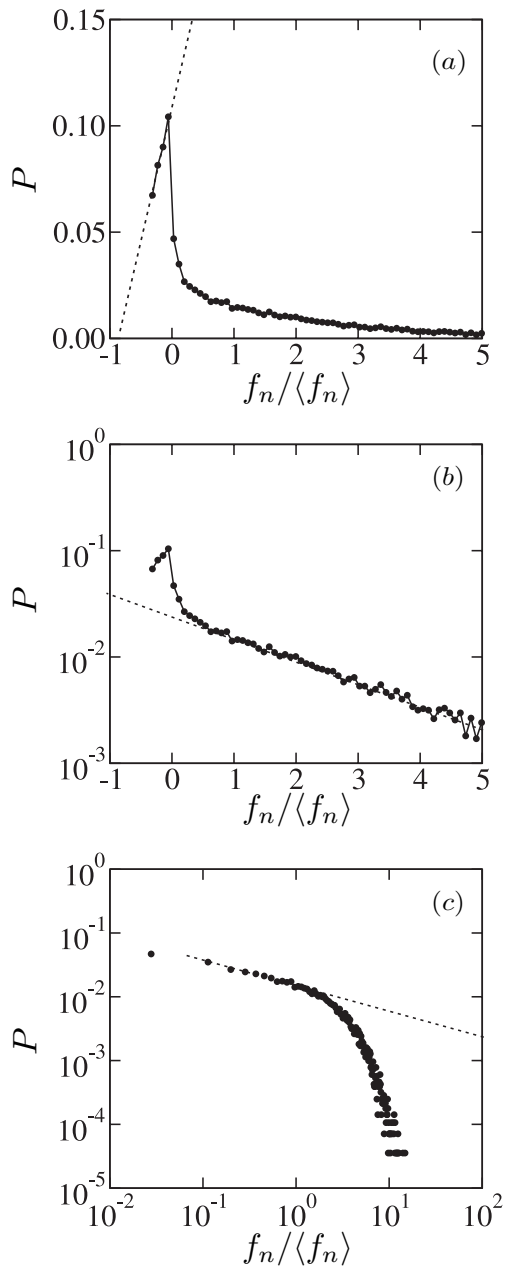}
\caption{Probalility distribution function $P$ of normal forces $f_n$ normalized by the mean $\langle  f_n \rangle$ represented on linear scale (a), log-linear scale (b) and log-log scale (c).\label{fig13}}
\end{figure}
 
The probability distribution function $P$ of normal bond forces is shown in Fig.~\ref{fig13}. The largest forces belong to the compressive network (involving no limiting threshold) whereas the tensile forces extend down to the capillary force threshold $-f_0$. The distribution reveals three different force intervals with different statistics in each interval:
\begin{equation} \label{eq:force_pdf}
P(f_n) \propto
\left \{
\begin{array}{ll}
e^{-\beta \frac{f_n}{\langle f_n  \rangle}}     &     {\rm for}\;    f _n>  \langle f_n  \rangle \\
{\left ( \frac{f_n}{\langle f_n  \rangle}\right )} ^{-\alpha}  &     {\rm for} \;      0<f_n<\langle f_n  \rangle \\ 
\gamma \frac{f_n+f_0}{\langle f_n  \rangle} + P_0 &     {\rm for} \;      -f_0<f_n<0 
\end{array}  \right.,
\end{equation} 
where $\langle f_n \rangle$ is the average normal force, $\alpha \simeq 0.4$, $\beta \simeq 0.16$, $\gamma \simeq 0.15$ and $P_0=P(-f_0)$. The distribution for compressive forces is reminiscent of that observed in dry granular media \cite{radjai99,Majmudar05,Mueth98}. The exponent $\beta$ has, however, a smaller value in the cohesive case. This means that tensile forces allow the packing to sustain stronger compressive force chains than in a cohesionless material. The power law  distribution of weak compressive forces (the range $0 < f_n < \langle f_n  \rangle$) over nearly one decade suggests that, in the presence of cohesive bonds, the contact network in this range is self-similar. These contacts are referred to as ``weak contacts'' and they play an important role in propping strong force chains in the complementary ``strong network'' (the range $f _n >  \langle f_n  \rangle$) \cite{radjai98b}. The probability 
distribution $P$ is robust and the exponents do not evolve during shear.  

\begin{figure}
\includegraphics[width=7cm]{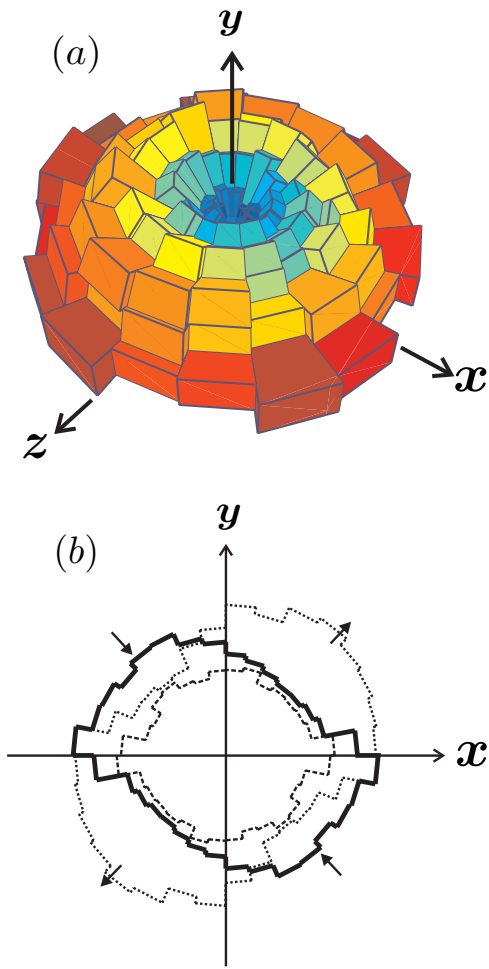}
\caption{(Color online) (a) Polar diagram of the probability distribution function of bond direction; (b) polar diagrams of strong (solid line), weak (dashed line) and tensile (dotted line) bonds along the vertical plane. The direction of extension and compression are represented (arrows).\label{fig14}}
\end{figure}

Figure \ref{fig14}(a) shows a polar diagram of the probability distribution function $P(\bm n)$ 
of bond directions $\bm n$ at the end of shearing for $w=0.02$ (the distribution being similar in other cases). The distribution is nearly isotropic along the shear plane ($xz$ plane in the figure) but it shows a pronounced anisotropy in the shear direction along the vertical plane ($xy$ plane in the figure). Fig.~\ref{fig14}(b) displays three separate polar diagrams of the strong, weak and tensile bonds along the vertical plane. The strong compressive bonds form an anisotropic distribution with its longest axis oriented at $45^\circ$ to the horizontal, as expected. The weak compressive bonds have a nearly isotropic distribution. Finally, the tensile bonds have an anisotropic distribution whose longest axis is perpendicular to that of strong compressive bonds. It has also been argued that the tensile forces play the same role in sustaining the strong force chains as the weak network \cite{Radjai00}. 

\section{\label{sec:analysis}A microscopic analysis}

In this section, we would like to propose an expression for the Coulomb cohesion $c$ as a function of parameters pertaining to the granular microstructure in the presence of liquid bonds. When the Mohr-Coulomb model (Eq.~\ref{eq:Mohr-Coulomb}) is valid also in the range of negative  stresses down to the tensile strength $-\sigma^t$, the Coulomb cohesion is related to the tensile strength by $c=\mu \sigma^t$. Keeping with this assumption, we therefore consider the tensile strength.
Theoretical evaluation of the tensile strength and its comparison with experiments or simulations can be found in recent literature concerning wet granular materials \cite{Pierrat97,Groger03}. The first theoretical expression of tensile strength was proposed by Rumpf  \cite{Rumpf70}. The key point is how capillary forces are mobilized in a wet material and what are the relevant structural parameters. Several parameters, such as the particle size and its distribution, the solid fraction $\phi$ and the internal coefficient of friction $\mu$, influence the macroscopic cohesion. The effect of the water content $w$ is somehow counter-intuitive. Indeed, the Coulomb cohesion saturates as the water content is increased while we expect that both the number of liquid bonds and the strength of each bond should increase with the water content and lead to higher cohesion. 

In order to estimate the tensile strength from contact forces, we consider the general expression of the stress tensor ${\bm \sigma}$ in a granular material. This is an average quantity with a well-established expression involving contact forces ${\bm f}^k$ and inter-center distances ${\bm \ell}^k$ \cite{christoffersen81,moreau97}:
\begin{equation} \label{eq:stress_tensor}
\sigma_{i j} = \frac{1}{V} \sum_{k \in V} f_i^k \ell_j^k,
\end{equation}
where $i$ and $j$ refer to components and $V$ is the control volume. The derivation of this expression is independent of the nature of interactions so that Eq.~\ref{eq:stress_tensor} holds also in the presence of capillary forces. In this case, the set of contact points is simply extended to cover 
capillary bridges. It can be shown that ${\bm \sigma}$ is symmetric if no torques are transmitted at the interaction points. 
From Eq.~\ref{eq:stress_tensor}, the stress $\sigma_{11}$ in the direction of extension is given by:
\begin{equation}
\sigma_{11}=n_w {\langle f_1\ \ell_1 \rangle}_b,
\label{eq:s11}
\end{equation} 
where $n_w$ is the number of bonds per unit volume, $f_1$ and $\ell_1$ are the components along the direction of extension. The symbol ${\langle \cdots \rangle}_b$ designs averaging over all bonds in the control volume $V$.
Let ${\bm n}$ and ${\bm t}$ be the normal unit vector and a tangential unit vector, respectively, at a given bond. Then, $f_1=f_n n_1+f_t t_1$ and $\ell_1=\ell n_1$ where $f_n$ and $f_t$ are the normal and tangential components of the force, and $\ell$ is the length of the branch vector ${\bm \ell}$. Let us set $n_1=\cos \Theta$ and $t_1=\sin \Theta$, where $\Theta$ is the angle between ${\bm n}$ and the direction of extension. Substituting in Eq.~\ref{eq:s11} and assuming for simplicity that $f_n$ and $f_t$ are independent of $\Theta$, we arrive at the following expression: 
\begin{equation}
\sigma_{11}=\frac{1}{2}\ n_w {\langle f_n\ \ell \rangle}_b.
\label{eq:s11-2}
\end{equation} 

In solid state physics, a theoretical tensile strength $\sigma^{th}$ is introduced from inter-atomic forces by assuming that the same failure threshold is reached simultaneously for all pairs of atoms in the direction of traction \cite{ElYous05}. In a similar approach, we may introduce a theoretical tensile strength for a wet particle assembly by replacing $f_n$ in Eq.~\ref{eq:s11-2} by the capillary force threshold $f_0$:
\begin{equation}
\sigma^{th} = \frac{1}{2}\ n_w {\langle f_0\ \ell \rangle}_b.
\label{eq:sigma-th1}
\end{equation}
The bond density $n_w$ is simply half the average number of bonds per particle divided by the free volume, i.e. the mean volume $V_p$ of a Voronoi cell surrounding the particle. The latter is simply the average particle volume $(1/6)\pi \langle d^3 \rangle$ divided by the solid fraction $\phi$. Introducing these expressions in Eq.~\ref{eq:sigma-th1} and using Eq.~\ref{eq:kappa}, we get
\begin{equation}
\sigma^{th} = \frac{3}{2\pi} \kappa\, \phi\, z \frac{\langle (R_1+R_2)\sqrt{R_1 R_2}) \rangle}{\langle d^3 \rangle} = \frac{3}{4\pi} s\frac{\kappa\, \phi\, z}{\langle d \rangle},
\label{eq:sigma-th2}
\end{equation}
where $z$ is the average number of bonds per particle and we have
\begin{equation}
s=\frac{\langle d^{1/2} \rangle \langle d \rangle \langle d^{3/2} \rangle} { \langle d^3 \rangle}.
\end{equation}
In derivation of the expression of $s$, it was assumed that the particle radii $R_1$ and $R_2$ are not correlated. This means that various granulometric classes are homogeneously distributed in the bulk. 
 It is easy to see that for a uniform size distribution, $s$ varies from $8/15$ to $1$ as the smallest particle size increases from $0$ to the mean particle size. For a monodisperse assembly, we have $s=1$ (see Table~\ref{tab2}). Eq.~\ref{eq:sigma-th2} is similar to the expression proposed first by Rumpf \cite{Rumpf70} for monodisperse materials (so, without the $s$ prefactor), and recently derived from the stress tensor by Gr\"{o}ger et al. \cite{Groger03}. Our equation~\ref{eq:sigma-th2} accounts in a simple way for polydispersity and the correlation between the capillary force threshold $f_0$ and the particle size $d$. For real polydisperse materials, the factor $s$ is crucial for comparing the model with experiments.   

By definition, the Coulomb cohesion is the yield shear stress at zero confining pressure in which case  the capillary forces are the only forces acting in the material. In this limit, the normal stress $\sigma$ on the shear plane is simply equal to the average capillary force divided by the sample section $S$. We have seen that ``gap'' liquid bonds (without contact) contribute only marginally to force transmission ($z^-$ being below one bond per particle); see Fig. \ref{fig12}. It is thus reasonable to assume that the capillary force at each bond is $f_0$. This means that, in the absence of confining stresses, we may 
set $\sigma=\sigma^{th}$. Then, the shear stress at yield is the theoretical cohesion $c^{th}$ given by 
\begin{equation}
c^{th}=\mu \sigma^{th} = \frac{3}{4\pi} \mu s  \frac{\kappa\, \phi\, z}{\langle d \rangle}.
\label{eq:cth}
\end{equation}

\begin{figure}
\includegraphics[width=7cm]{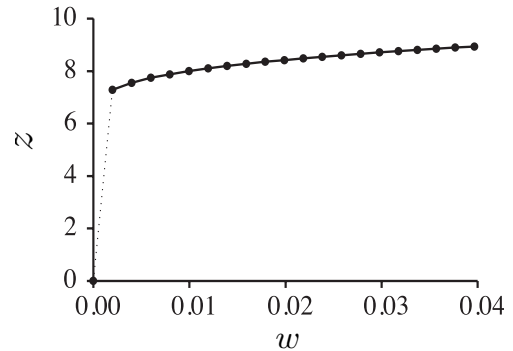}
\caption{The initial wet coordination number in simulations as a function of water content.\label{fig16}}
\end{figure}

It is important to note that the water content does not enter the above expression of $c^{th}$. The only parameter related to water is $\kappa$. This suggests that the water content manifests itself mainly through the wet coordination number $z=z^- + z^{c-} + z^{c+}$. In particular, the cohesion $c_m$ at saturation corresponds to the saturation of $z$ as the water content $w$ is increased. In fact, when a certain amount of water is homogeneously distributed in the whole sample within the de-bonding distance, one finds that $z$ increases with $w$ and saturates beyond $w=w_m$. This is shown in Fig.~\ref{fig16} for the initial configuration of our numerical samples. 
This is a purely geometrical effect related to steric exclusions among particles and it 
explains therefore the saturation of cohesion with water content according to Eq.~\ref{eq:cth}.

Although this geometrical saturation should dominate in the pendular state, it does not elude that other mechanisms might play a role in the experiments. In particular, as the liquid content is increased, the liquid bonds may coalesce at least locally, leading to bond saturation. We did not observe liquid bond clustering at the visible parts of our samples. 

\begin{figure}
\includegraphics[width=7cm]{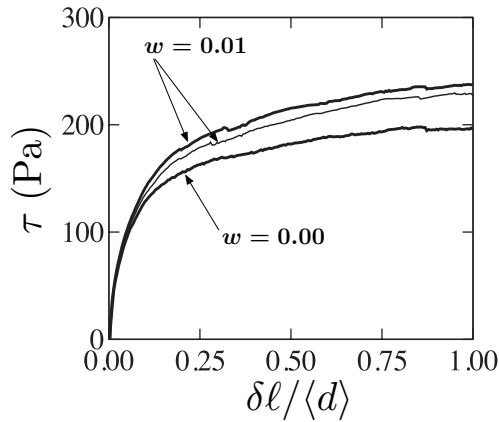}
\caption{The shear stress $\tau$ as a function of shearing distance $\delta \ell$ normalized by the average particle diameter $\langle d \rangle$ for a dry and two wet samples with different numbers of liquid bonds; see text. The confining stress is $\sigma = 300$ Pa.\label{fig17}}
\end{figure}

The initial value of $z$ depends also on the preparation protocol. Since $f_0$ is independent of $w$, the same amount of water can be distributed in such a way as to produce a lower number of liquid bonds and thus a lower macroscopic cohesion. This effect is illustrated in Fig.~\ref{fig17} where the stress-strain plots are shown for two initially identical configurations differing only in the number of liquid bonds for the same water content. The simulation was carried out by using the second method of liquid redistribution (section \ref{sec:simu}). In the sample where half of the water bonds has been removed, the wet coordination number increases with deformation. But in the initial stages of deformation, the cohesion is close to half that of the sample involving a double number of water bonds, and it increases as the wet coordination number grows. 
 
We summarize in Table \ref{tab2} the theoretical estimates $c^{th}$ and the measured values $c_m$ of the saturated Coulomb cohesion  for all our experimental and numerical samples together with the values of the parameters involved in Eq.~\ref{eq:cth}. The value of the polydispersity factor $s$ was calculated from the knowledge of the particle size distributions.   
Note that $c^{th}$ is in excellent agreement with $c_m$ both for experiments and simulations. It is noteworthy that if the prefactor $s$ were not incorporated in Eq.~\ref{eq:cth} (i.e. if Rumpf's equation had been used), the measured value of $c_m$ would be below $c^{th}$ by a factor $s$. 

This agreement between the theoretical estimate and experiments with sand and glass beads was obtained with $z = 6$ which is a reasonable value of the bond coordination number in the case of a homogeneous distribution of water in the bulk, and it is suggested by recent experimental observations \cite{Kohonen04,Fournier05}. Note, however, that in the case of fine-grain samples (sand and GB1 in Table \ref{tab2}) a closer agreement can be obtained with a lower value of $z$. This is suggestive in the sense that  
liquid bond clustering might indeed occur more frequently for fine grains and reduce thus the effective bond coordination number.

\begin{table}
\caption{Measured and theoretical parameters for all our experimental and numerical samples. The approximate value of  the bond coordination number $z$ for the experiments (indicated by a question mark) was suggested by the literature \cite{Kohonen04,Fournier05}.\label{tab2}}
\begin{ruledtabular}
\begin{tabular}{llllll}
 & Sand & GB1\footnote{tigthly-graded polydisperse glass beads} 
 & GB2\footnote{well-graded polydisperse glass beads} 
 & GB3\footnote{monodisperse glass beads} & Simulations \\
 \cline{2-6}
$\langle d \rangle$ (mm) & 0.16 & 0.45 & 0.60 & 1.00 & 1.65 \\
$s$                &  0.50 & 0.99 & 0.91 & 1.00 & 0.79 \\
$z$                & 6(?)  & 6(?) & 6(?)  & 6(?) & 9 \\
$\phi$            & 0.6    & 0.6  & 0.6   & 0.6 & 0.6\\
$\mu$            & 0.66  & 0.58 & 0.58 & 0.46 & 0.48 \\
$c_m$ (Pa)   & 600   & 350 & 300  & 150 & 120 \\
$c^{th}$ (Pa) & 709  & 438  & 302  & 158 & 118 \\
\end{tabular}
\end{ruledtabular}
\end{table}

By construction, the theoretical estimate is an upper bound for cohesion. For brittle materials, failure is initiated by the breakdown of a few bonds and propagates subsequently into the material. When this mechanism works, the tensile strength is not controlled by the average stress (as assumed in the derivation of $\sigma^{th}$) but by the largest local stresses, and hence the effective strength is far below the theoretical one (in proportion to the stress concentration factor) \cite{Herrmann90}. Hence, the nice agreement of the theoretical estimate both with simulations and experiments suggests that the failure of our wetted granular materials  is ductile and the shear strength is controlled by the mean tensile force.

\section{Conclusion}

We performed experiments and discrete element simulations to analyze the Coulomb cohesion of wet granular media in the pendular state. It was shown that the Coulomb cohesion increases with water content and saturates to a maximum value that depends only on the nature of the material. An interesting aspect that was partly investigated by simulations is that the cohesion is basically controlled by the number of liquid bonds.
This suggests that the saturation of Coulomb cohesion occurs since new bonds are hardly formed beyond a certain amount of the water content. On the other hand, the capillary failure threshold is nearly independent of the local liquid volume.

Starting with the expression of the stress tensor, we also introduced a novel expression for the Coulomb cohesion as a function of material and structural parameters. This expression extends the classical model of Rumpf  to polydisperse materials. We found that our model is in excellent agreement with experimental and numerical data. 

From numerical data, we analyzed the connectivity and anisotropy of different classes of liquid bonds according to the sign and level of the normal force as well as the bond direction. We found that weak compressive bonds are almost isotropically distributed whereas strong compressive and tensile bonds have a pronounced anisotropy. It was shown that the probability distribution function of normal forces is exponentially decreasing for strong compressive bonds, a decreasing power-law function over nearly one decade for weak compressive bonds and an increasing linear function in the range of tensile bonds.  

In the extension of this work, it is essential to evaluate the limits of the model by considering other materials and non monotonous loading paths. In particular, we would like to study the shear strength of granular media with a larger polydispersity than materials that were used in the present investigation. Since the distribution of liquid bonds seems to be a major parameter for the cohesion, it also merits to be investigated in more detail experimentally. Finally, an interesting application of the ideas put forward in this paper would be to examine by which mechanisms the cohesion of a sample of wet sand increases as a result of compactification. 

\begin{acknowledgments}

We gratefully acknowledge J.-Y. Delenne, R. Peyroux and F. Souli\'e for interesting discussions.
The first author would like to thank DGA (the french ministry of defense) for providing financial support for his PhD work.

\end{acknowledgments}

\newpage
\bibliography{references}

\end{document}